\documentclass[apj]{emulateapj}
\journalinfo{\textsc{The Astrophysical Journal}}
\slugcomment{Received May 3, 2021; Revised December 30, 2021; Accepted January 14, 2022}
\usepackage{graphicx}
\usepackage{lineno}
\usepackage{amsmath}

\usepackage[colorlinks,
        hypertexnames=false,
        pagebackref=false,
        linkcolor=blue,    
    citecolor=blue,        
    filecolor=magenta,     
    urlcolor=blue          
]{hyperref}                

\newcommand{\eg}{\textit{e.g.,}}

\newcommand{\degree}{^\circ}

\newcommand{\kms}{km s$^{-1}$} 
\newcommand{\mss}{m s$^{-2}$}  
\newcommand{\Rsun}{R_{\odot}}  

\shorttitle{Eruption of the EUV hot-channel by tether-cutting reconnection}
\shortauthors{Vemareddy et al.}

\begin{document}

\title{Eruption of EUV Hot-Channel near Solar Limb and Associated Moving Type-IV Radio Burst}


\author{P.~Vemareddy}    
\affil{Indian Institute of Astrophysics, II Block, Koramangala, Bengaluru-560 034, India}

\author{P.~D\'emoulin}     
\affil{LESIA, Observatoire de Paris, Universit\'e PSL, CNRS, Sorbonne Universit\'e, Universit\'e de Paris, 5 place Jules Janssen, F-92190 Meudon, France}
\affil{Laboratoire Cogitamus}

\author{K.~Sasikumar~Raja}   
\affil{Indian Institute of Astrophysics, II Block, Koramangala, Bengaluru-560 034, India}

\author{J.~Zhang}   
\affil{Department of Physics and Astronomy, George Mason University, Fairfax, VA 22030, USA}

\author{N.~Gopalswamy}   
\affil{Goddard Space Flight Center, Greenbelt, USA}

\author{N.~Vasantharaju}   
\affil{Indian Institute of Astrophysics, II Block, Koramangala, Bengaluru-560 034, India}
\affil{Dipartimento di Fisica e Astronomia “Ettore Majorana”—Sezione Astrofisica, Università degli Studi di Catania, Via S. Sofia 78, I-95123 Catania, Italy}

\begin{abstract}
Using the observations from Solar Dynamics Observatory, we study an eruption of a hot-channel flux rope (FR) near the solar-limb on February 9, 2015. The pre-eruptive structure is visible mainly in EUV 131\,\AA\ images with two highly-sheared loop structures. They undergo slow rise motion and then reconnect to form an eruptive hot-channel as in the tether-cutting reconnection model. The J-shaped flare-ribbons trace the footpoint of the FR which is identified as the hot-channel. Initially, the hot channel is observed to rise slowly at 40 kms$^{-1}$, followed by an exponential rise from 22:55 UT at a coronal height of 87$\pm$2 Mm. Following the onset of the eruption at 23:00 UT, the flare-reconnection adds to the acceleration process of the CME within 3\,$\Rsun$. Later on, the CME continues to accelerate at 8 \mss\ during its propagation period. Further, the eruption launched type-II followed by III, IVm radio bursts. The start and end times of type-IVm correspond to the CME core height of 1.5 and 6.1 $\Rsun$, respectively. Also the spectral index is negative suggesting the non-thermal electrons trapped in the closed loop structure. Accompanied with type-IVm, this event is unique in the sense that the flare ribbons are very clearly observed along with the erupting hot channel, which strongly supports that the hooked-part of J-shaped flare ribbons outlines the boundary of the erupting FR.   
\end{abstract}


\section{Introduction}
\label{Intro}
Magnetic reconnection is a fundamental physical process that has a prime role especially in releasing magnetic energy during solar eruptions \citep[\eg ][]{Priest2000,Pontin2012}.  When a magnetic configuration has an excess of magnetic energy, a trigger is required to initiate the eruption. Several physics-based models have been invoked to explain the onset mechanism of an eruption. Of these, the tether-cutting or flux cancellation model \citep{Ballegooijen1989, Moore2001} and the magnetic breakout model \citep{Antiochos1999} come under the category of reconnection based models. In these models, the initial sheared core field along the magnetic polarity inversion line (PIL) is enveloped by the overlying potential arcade and the reconnection either in the inner arcade or in the overlying field triggers the eruption. Other models rely on an ideal MHD instability, viz., kink and torus instabilities, with the presumption that the pre-eruptive structure is a twisted magnetic flux rope \citep[FR,][]{Forbes1991, Torok2005, Kliem2006, Forbes2006}. The destabilization is realized when either a critical twist or a too steeply decreasing background field is reached. Although the mechanism triggering an eruption is different in the above models, magnetic reconnection induced underneath the uplifting FR plays an important role during the eruption since it allows to have a successful ejection into the heliosphere by transforming the stabilizing upper arcade to the twisted field surrounding the original erupting core field \citep{Lin2000, Aulanier2010, Welsch2018}. The resulting enlarged FR is typically observed as a coronal mass ejection (CME) and the remaining reconnected field below as post-flare arcades rooted in flare ribbons.

The high resolution space-based observations from the Transition Region and Coronal Explorer \citep{Handy1999} and the Atmospheric Imaging Assembly (AIA; \citealt{Lemen2012}) on board the Solar Dynamics Observatory (SDO; \citealt{Pesnell2012}) provide key observations to understand the initiation mechanism. Also H$\alpha$ filaments, prominences, and X-ray sigmoids are precursor features of an eruption. These features have been interpreted in the same physical framework with the presence of a FR, or alternately, of a sheared magnetic arcade \citep[\eg ][]{Gibson2006, Vemareddy2014, Green2011}. Recent studies discovered that the FR configurations prevailed with hot coronal conditions of several million degree kelvin temperatures visible in EUV hot channels \citep{ZhangJie2012, ChengX2013}. 

The tether-cutting reconnection \citep{Moore2001, Moore2006} was found to play both a formation and triggering role. During the eruption it evolves to the run-away tether-cutting reconnection below the sheared arcade, and finally the eruption of the core field occurs and a CME is launched \citep{Yurchyshyn2006, LiuR2010, Vemareddy2012}. A reconnection in the overarching loop structure is identified mostly as a remote brightening following the onset of this eruption. The trigger could also be due to the dynamics of the overarching loops with exceeding helical twist as suggested by the EUV observations of the sigmoidal structure \citep{Vemareddy2014}.

Several observational reports indicated that magnetic cancellation at the photospheric level, induce tether-cutting reconnection between two sets of highly sheared magnetic arcades over hours before eruption.  This implies the formation of a FR which later erupts \citep{Green2011, Vemareddy2017, XueZhike2017}. Further, the reconnecting sheared arcades are sometimes identified as two lobes of a sigmoid. Using high-resolution Interface Region Imaging Spectrometer (IRIS) observations, \citet{ChenHuadong2016} reported that the tether-cutting reconnection occurs between the sheared magnetic configurations of two filaments. This leads to a flare-associated CME. The tether-cutting reconnection may also be responsible for the trigger of the CME eruption in the complex double-decker filament channel. The involved magnetic configuration was proposed to be either a double FR configuration or a single FR on top of sheared arcade \citep{LiuR2012, Vemareddy2012, Joshi2020_SeqLidRemov}. 

In this article, we report a unique observation of the eruption of a hot channel FR located near the solar limb.  These observations have different projection effects than similar events observed on the solar disk. This allows us to differentiate the coronal structures involved, which is crucial for understanding the triggering mechanisms of the eruptions \citep{ChenHuadong2016, Vemareddy2017}. 
In addition, this eruption triggers radio bursts of type III, II, IV. Flare reconnection converts a part of the magnetic energy to accelerate electrons along magnetic field lines. These fast moving electrons set-up plasma oscillations (called Langmuir waves) in the solar corona during their passage and their subsequent conversion into electromagnetic waves produces radio emission as type III burst \citep{Ginzburg1958, Zheleznyakov1970, Melrose1980, Sasikumar2013, Day2019, Ndacyayisenga2020}. Type II are slow drifting bursts ($\approx 1000 km~s^{-1}$) generated by non-thermal electrons accelerated at shocks propagating through the solar corona and interplanetary medium (IPM) \citep{Payne-Scott1947, Nel1985, Nindos2008, Cho2013}. 
Moving type-IV radio bursts \citep{Ste1982, Leb2000, Sasikumar2014} are generally believed to be produced by electrons trapped within the erupting closed structures, 
which are proposed to be the radio signatures of the hot channel FRs as evidenced by the radio imaging observations \citep{Demoulin2012, WuZhao2016, Vasanth2016}. 

We describe the EUV and magnetic observational data of the erupting hot channel in Section~\ref{Obs}. Then, in Section~\ref{Res} we first analyse the onset of the eruption, second we study the kinematics of the eruption up to 30 $\Rsun$ to constrain the physical mechanisms involved, and finally we explored the link between the radio bursts and the eruption. A summarized discussion is given in Section~\ref{Summ}. 


\begin{figure*}[!ht] 
	\centering
	\includegraphics[width=.99\textwidth,clip=]{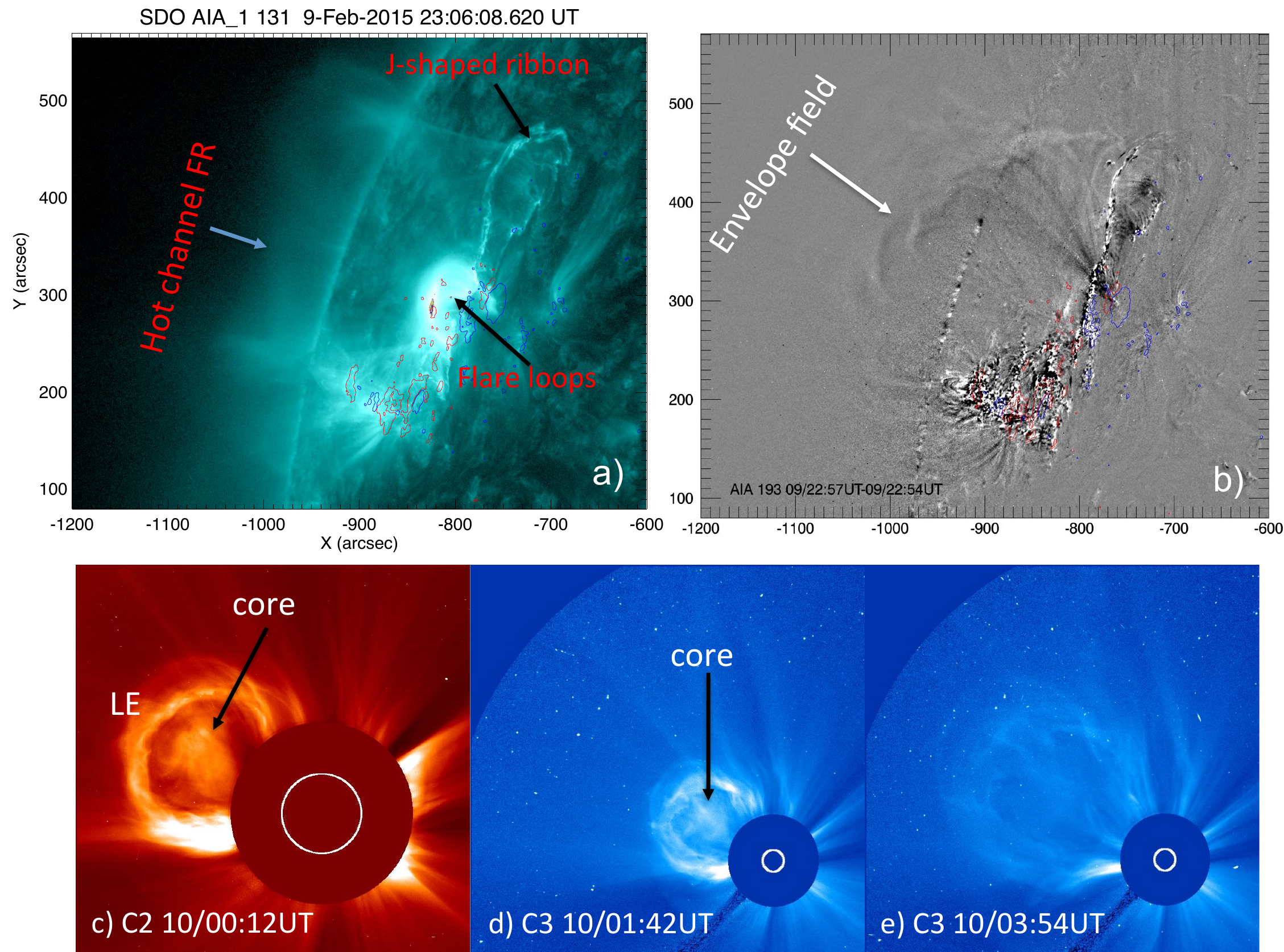}
	\caption{Hot-Channel eruption from the near limb AR 12282. {\bf a-b)} AIA images taken during the flare impulsive phase. A diffused coherent structure (marked hot channel/FR) is visible in AIA 131~\AA\ which we refer to as FR. A difference image of AIA 193~\AA\ shows the morphology of the envelope structure above the AR. Contours of the co-aligned LOS magnetic field, $\pm 90$ G (red/blue), are over plotted in order to locate the magnetic origins of the erupting structure. {\bf c-e)} White-light observations of the CME from LASCO C2/C3 showing distinct leading edge, cavity and core part. The core is getting less contrasted as the CME expands further in C3 field-of-view. } 
	\label{Fig_aia_CME}
\end{figure*}

\begin{figure*}[!ht] 
	\centering
	\includegraphics[width=.99\textwidth,clip=]{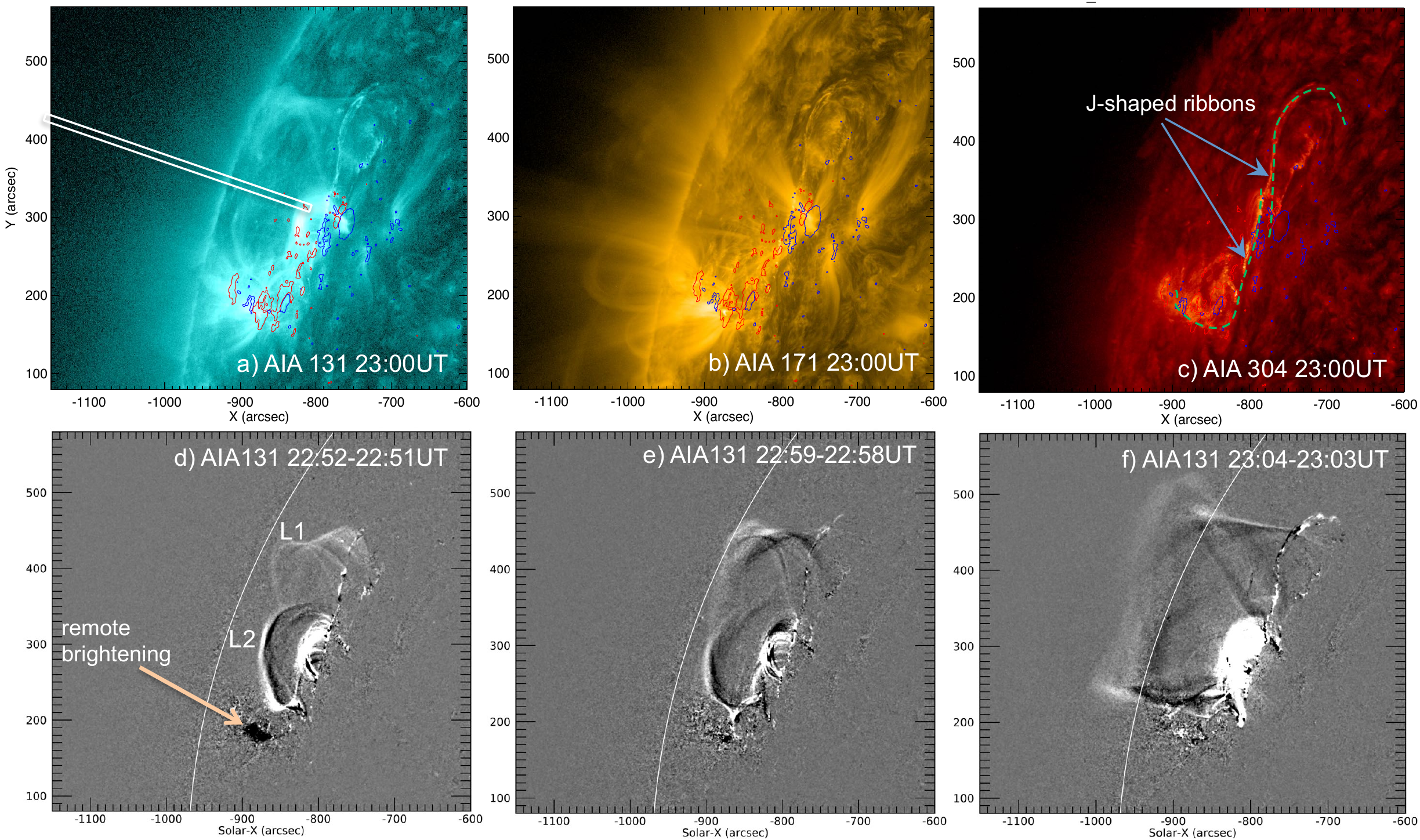}
	\caption{ Onset of the eruption from AR 12282. {\bf a-c)} Images of the erupting region in AIA 131, 171, 304~\AA\ channels. J-shaped ribbons are observed in AIA 304~\AA\ from 22:55 UT onward. Contours $\pm90$\,G (red/blue) of LOS magnetic field are also overlaid.  For kinematic analysis, slit position is shown in panel a . {\bf d-f)} difference images of AIA 131~\AA\ during the onset of the eruption. Two nearby loop structures L1 and L2 first rise up, then reconnect at a coronal cross point to form a single larger loop structure and flare loops below. A remote brightening location is marked with an arrow. The white curve outlines the edge of the limb. This scenario of onset of the eruption is more comprehended with an online animation accompanied with this figure. It is prepared from sequence of AIA 131~\AA~images (left panel) and the running difference images of AIA 131~\AA~(right panel). Animation start (end) time is 22:30 (23:18) UT.   
	} 
	\label{Fig_diff}
\end{figure*}
 
\begin{figure*}
	\centering
	\includegraphics[width=0.99\textwidth,clip=]{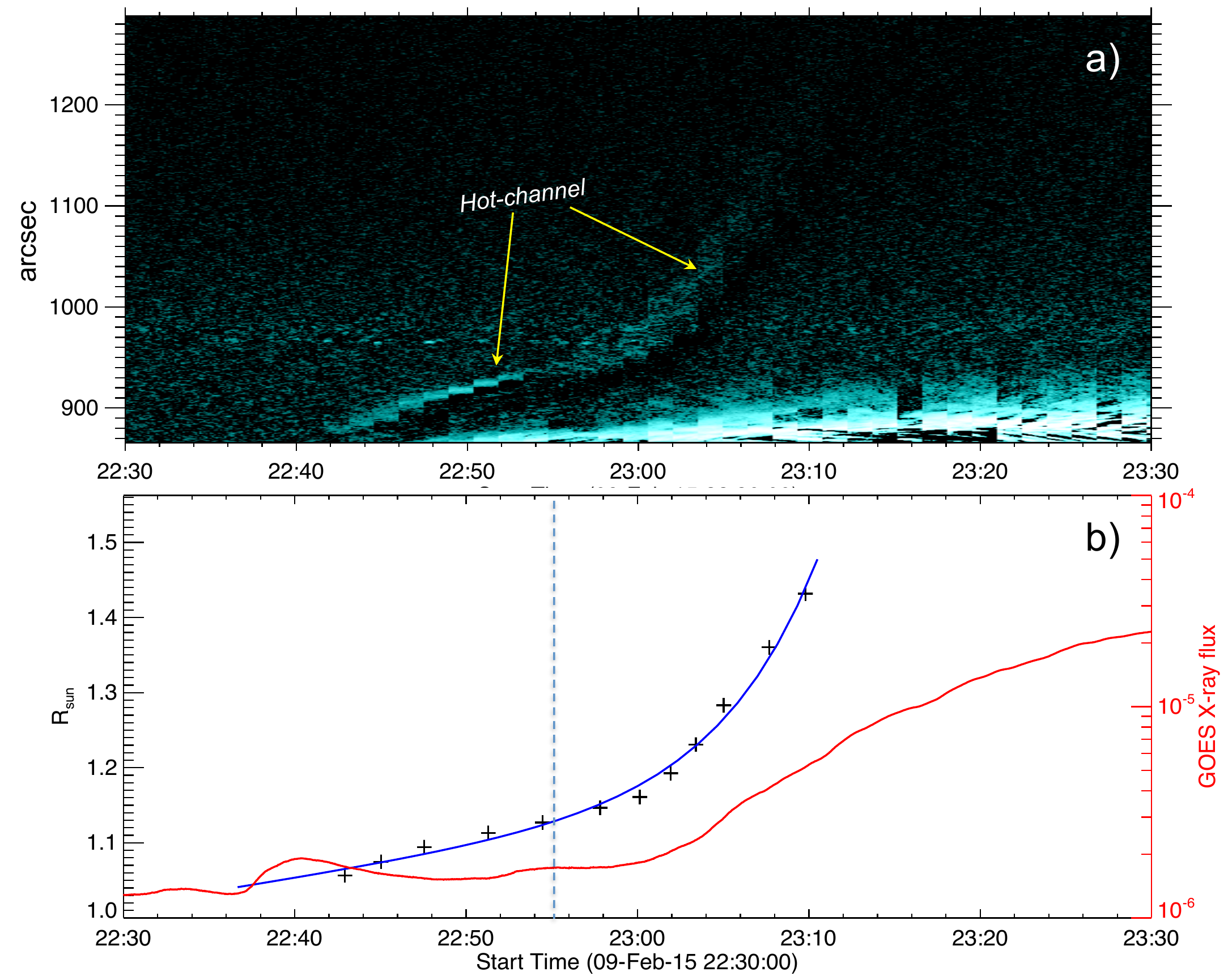}
	\caption{ Upward rise motion of the hot-channel with time. {\bf top:} space-time map prepared from the slit as shown in Figure~\ref{Fig_diff}a. Trace of the hot channel is indicated with arrow. {\bf bottom:} height-time plot of the hot-channel derived from the space-time map. Note that the height is corrected for projection effect such that the motion is radially outward. The data points along the hot-channel trace are fitted with linear-cum-exponential function (blue curve) and the blue vertical dashed line marks the time at which the linear velocity is dominated by the exponential growth of the velocity and is referred as critical time for the eruption onset. The corresponding height is referred as critical height which in this case is 0.126 R$_\odot$ (87 \,Mm.). GOES X-ray flux is also shown in red curve with y-axis scale on right side. Noticeably, the eventual eruption accompanied by flare occurs at 23:00 UT. }
	\label{fig_aia_ht}
\end{figure*} 
 
\begin{figure*}[!ht] 
\centering
\includegraphics[width=.99\textwidth,clip=]{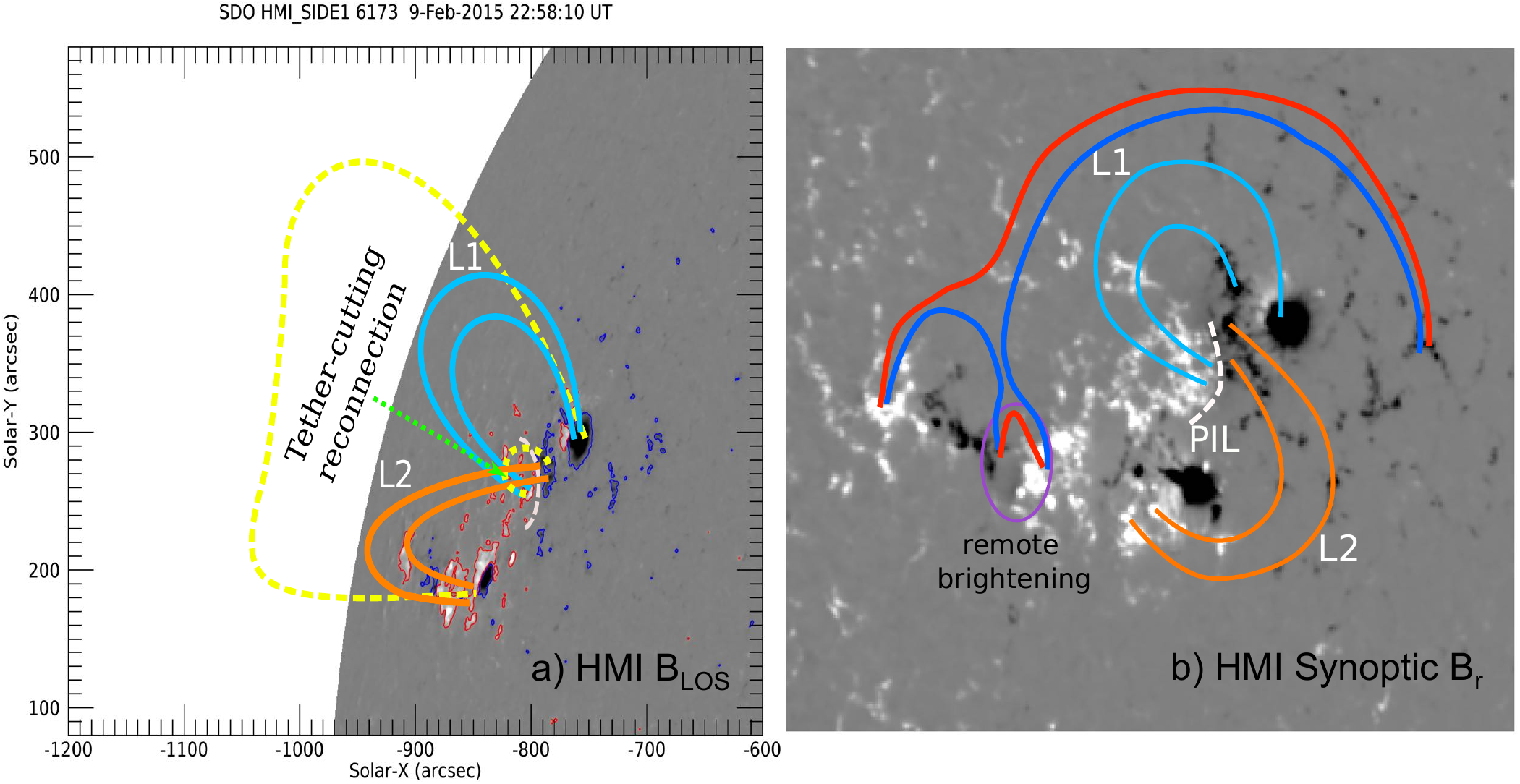}
\caption{ Schematic illustration of the possible magnetic structure undergoing the tether-cutting reconnection. {\bf a)} HMI LOS magnetogram displaying the distribution of the magnetic field. Contours of $\pm$90 G are overlaid. The loop systems L1 (cyan) and L2 (orange) are sketched to reveal their magnetic origins. White-dashed curve is the PIL. Dashed yellow curves refer to the reconnection bi-product of the loop systems L1 and L2
{\bf b)} HMI synoptic radial field distribution on February 11 at 12:00 UT. The map is the result of cylindrical equal area projection onto the disk center. White dashed curve refers to the PIL with a possible sheared arcade all along. The loop systems L1 (cyan) and L2 (orange) are sketched as if they were seen from above (dashed yellow curves for the reconnection bi-product are not shown). In addition, the purple oval is the location of the remote brightening with converging opposite magnetic polarities. The reconnection scenario causing the remote brightening is also sketched in blue (before reconnection) and red (after reconnection).  
}
\label{Fig_schem}
\end{figure*}
\section{Overview of the Observations}
\label{Obs}

The studied eruption event is well captured in the EUV observations of AIA/SDO and the white-light images from the Large Angle and Spectrometric Coronagraph (LASCO, \citealt{Brueckner1995}) on board the \textit{Solar and Heliospheric Observatory} (SoHO). The eruption originated from AR 12282 on February 9, 2015. The AR was located near eastern limb (N15$\degree$ E60$\degree$) it is easy to study the connection between the CME manifestation in the LASCO field-of-view (FOV) with the eruption features on the disk (Figure~\ref{Fig_aia_CME}). 

The eruption is initiated at 22:42 UT on February 9, 2015 with two loop structures (in the form of flux threads) crossing each other in projection on one side (shown at 22:52 UT in Figure~\ref{Fig_diff}d). After 22:55 UT, a single loop structure rises during the early phase of the eruption observed in AIA 131~\AA\ (Figures~\ref{Fig_aia_CME}a,~\ref{Fig_diff}d). This structure appears diffuse in AIA 131 and 94~\AA\ channels while it is not present in other AIA passbands (Figure~\ref{Fig_diff}a-c). Therefore, this erupting feature is a hot channel as observed in other events \citep[\eg  ][]{ZhangJie2012}. From AIA 193~\AA\ observations (Figure~\ref{Fig_aia_CME}), the plasma loop morphology reveals the envelop field extended up to 1.3\,$\Rsun$. 

The magnetic origins of this erupting feature is disclosed by the line-of-sight (LOS) magnetic field observations obtained from the \textit{Helioseismic Magnetic Imager} (HMI; \citealt{schou2012}) on board SDO. The AR belongs to the $\beta$-class with a quadrupolar magnetic configuration having two sunspots of negative polarity and a following dispersed positive polarity (Figure~\ref{Fig_schem}). The polarity inversion line (PIL) of the western bipole is with strongly sheared field.

Subsequent rising motion of the erupting feature manifests as a CME, which was first entering into C2 FOV at 23:24 UT on February 9 and C3 FOV at 00:06 UT on February 10. Representative images of the CME are displayed in the bottom panels of Figure~\ref{Fig_aia_CME}. Three part structures of the CME, i.e., leading edge, dark cavity and bright core, are well observed in C2 images \citep{Illing1985,Vourlidas2013, Vourlidas2020}. These observations are typically interpreted with the presence of a FR with/without a prominence/filament embedded in its core. However, the presence of the core may not always require an eruptive filament/prominence \citep{HowardTA2017} given the projection effects. So the plane-of-sky oriented flux rope in AIA images is not resolvable in the coronagraph images which although provides an impression of rotated flux rope in the core of the CME. After 01:42 UT on February 10, the white-light intensity reduces such that the CME appears as a diffused structure. 

This CME is accompanied by a GOES X-ray M2.4 class flare and type-II, III, IV radio bursts. 
We use the observations by Learmonth radio spectrometer located at North-West Cape, Western Australia \citep{Lobzin2010} and Wind/Waves instrument from Lagrangian point L1 \citep{Bougeret1995}. The former instrument operates in the frequency range of 25 - 180 MHz and the later observes in 20 KHz - 14 MHz.  

\section{Results}
\label{Res}
\subsection{Initiation of the Eruption: Observations}
\label{Onset}

The erupting feature is better visible in 131~\AA\ pass band compared to 94~\AA. Other channels exhibits no signatures of the erupting feature. The {initiation} is noticed with a loop system L1 as a small arc at 22:42 UT in 131~\AA, which is later on accompanied by a second loop L2 (defined in Figure~\ref{Fig_diff}d). The loops have one of their legs located nearby in the opposite polarities on both sides of the main PIL. The eruption onset is accompanied by two flare ribbons and flare loops in AIA 171 and 304~\AA\ images as displayed in the top row of Figure~\ref{Fig_diff}. 


Since the erupting structure is diffused, we studied time difference images of AIA 131~\AA\ to enhance the contrast from the background. Examples of difference images are displayed in bottom panels of Figure~\ref{Fig_diff} (also see the accompanied movie). In these images, the ascending loops and the underlying flare loops are clearly visible. L1 and L2 are seen distinctly until 22:55 UT. In the later evolution, these loops merged as a single structure without connection to the inner bipolar region.  This implies a coronal reconnection of the loops L1 and L2 which are then transformed to flare loops and a single larger structure above. We refer to this erupting structure as a hot-channel FR (HFR), with more justifying evidences below. The HFR ascends further and eventually erupts at around 23:00 UT (more details in Section~\ref{Kinematics}).  
 
At around 23:00 UT, the AIA 304~\AA\ images present two J-shaped flare ribbons with the approximately straight parts being close to the PIL (Figure~\ref{Fig_diff}c). Taken together the ribbons define a global S-shape which indicates that the HFR has a positive twist.  The straight parts of the ribbons are linked with flare loops underneath the rising HFR,  
and they become prominent as flare reconnection progresses. The hook part of the ribbons are surrounding the footpoints of the HFR as shown by the topological analysis of previous FR configurations \citep[e.g.][]{Demoulin1996JGR,titov2007,Savcheva2012}.
Said differently, the curved part of the ribbons provide the boarder of the FR footpoints. 

As the reconnection proceeds, the FR is growing in magnetic flux and ribbons are separating with a broader J-shape.  These ribbons are expected to be present at the base of quasi separatrix layers which are regions where the field line connectivity change drastically, where current layers are formed and where magnetic reconnection occurs \citep{Demoulin1996AA, Aulanier2010, Vemareddy2014_null, Zhao2016, Vemareddy2021_qsl}. Then, the emissions, observed in 131 \AA\ and moving away from the AR, are interpreted as the new reconnected field lines wrapped around the erupting FR. Furthermore, the northern hook is the clearest observed all along the event.  More than one hour before any eruption sign, plasma motions are already present all along this hook (see the attached movie in 131 \AA ). This implies that the FR was already present well before the eruption and weak reconnection was already active to heat up and to displace the plasma at the FR boarder.
We conclude that the EUV observations of the hot-channel structure together with J-shaped flare ribbons provide evidences for the magnetic FR before and during the course of the eruption.  To our knowledge, such observations of the hot channel with a clear link to the hook-shaped flare ribbon have not been shown previously. 
	
For a more clear picture of the onset of the eruption, we prepared space-time map of a slit placed across the hot-channel, as shown in Figure~\ref{Fig_diff}a. The map is displayed in Figure~\ref{fig_aia_ht}a, from it is clear that the rise motion starts at 22:42 UT, however at this time, one loop (L2) intersects the slit. As the rise motion progresses, several continuous loops form as delineated in the images after 22:55 UT. Around 23:00 UT, the two loops reconnect to form a single loop structure and then a rapid rise is evident from then onwards. 

\begin{figure*}[!ht] 
	\centering
	\includegraphics[width=.8\textwidth]{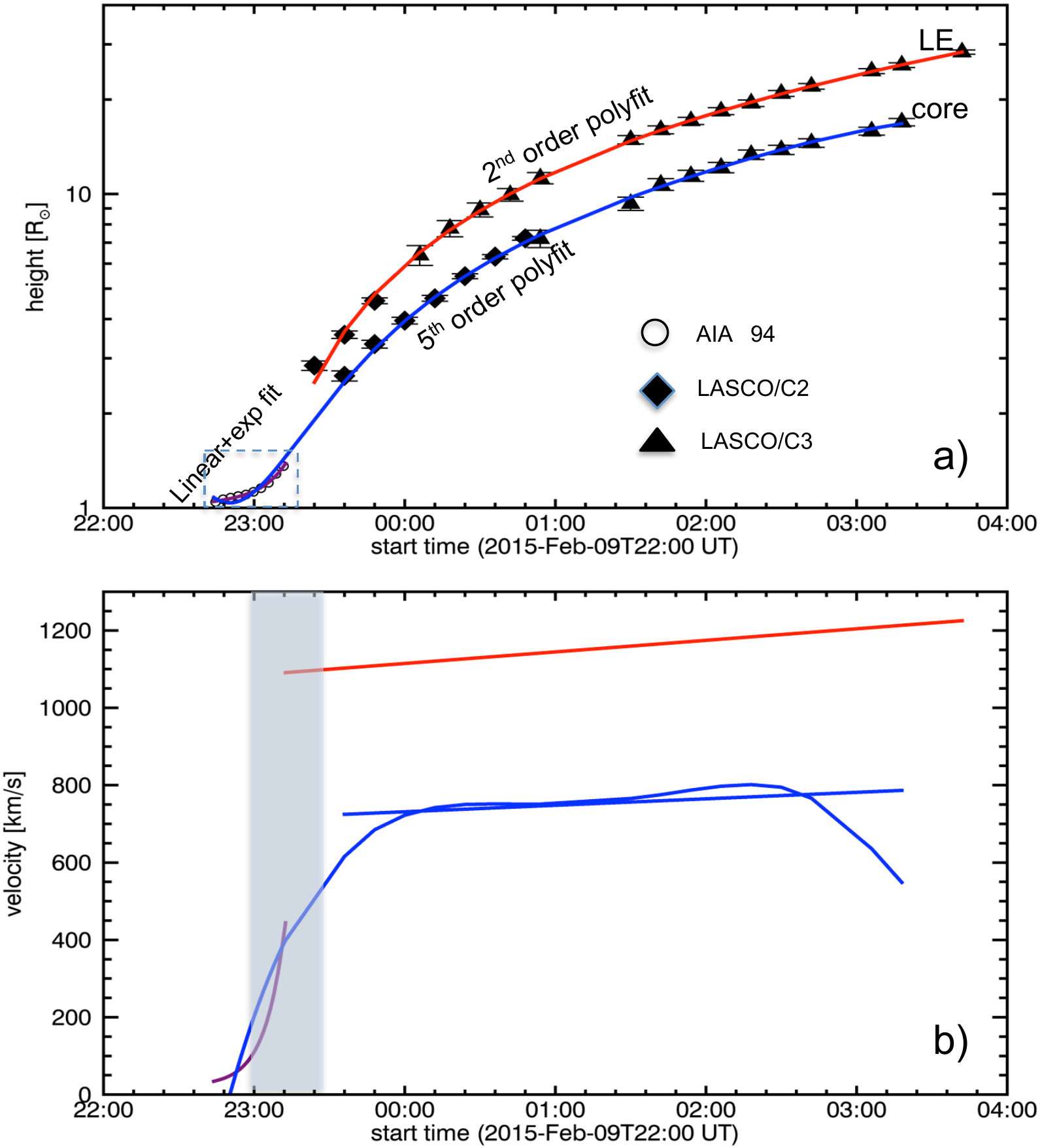}
	\caption{ Kinematics of the eruption event.
	{\bf a)} Height-time plot of the erupting HFR in AIA FOV (circle symbol), the LE and core in LASCO FOV. The purple and red-curves are the model fits to the height-time data with a linear-cum-exponential terms in AIA FOV and with 2nd order polynomial model in the LASCO FOV. The blue curve is a 5th order polynomial fit to the combined data of HFR and CME core, which fairly represents the data except in the AIA FOV. 
	{\bf b)} The derived velocity from the model fits to the height-time data. Blue curve is from 5th order polynomial fit; it presents some oscillatory behavior after 00:00 UT. The purple and red curves are linear-cum-exponential and 2nd order polynomial fit, respectively. The shaded region represents the rapid acceleration phase to the extent of peak flare time 09/23:35 UT, which is within 3 $\Rsun$. Within the AIA FOV, the velocity of the HFR rise motion increases from an initial velocity of 40 km\,s$^{-1}$ to 400 km\,s$^{-1}$. } 
	\label{Fig_ht_vel}
\end{figure*}

In Figure~\ref{fig_aia_ht}b, we plot the height-time observations of the hot-channel trace. Being at 60$^o$ longitude, the derived height-time data is corrected to compensate the projection effect by assuming a radial motion. We then fit this height-time data with a linear-cum-exponential model
\begin{equation} 
h(t) = C_0 + C_1 \, t + C_2 \, e^{t/\tau}
\label{eq_h(t)}
\end{equation}
where $C_0$, $C_1$, $C_2$, $\tau$ are the 4 free parameters of the fit \citep{ChengX2013}. This model accommodates the slow and rapid rise motions, so fitting very well to the data points (purple curve in Figure~\ref{Fig_ht_vel}). The growth time is $\tau = 24.87 $ minutes. This model fit allows to determine the transition time between the slow (nearly constant velocity) and the exponential growth stages. We estimate the critical time  $T_c=\tau \, ln(C_1 \, \tau/C_2)$ at which the exponential component of the velocity equals to the linear component. This is found to be February 09 at 22:55 UT which corresponds to a height of 1.126 $\Rsun$ (87 Mm). After this time, the exponential term in Equation~(\ref{eq_h(t)}) dominates and this behavior characterizes an instability. Note that at time rise motion also coincides with the disappearance of two loops to continuous loop structure, which is regarded as the flux rope. However, soft X-ray flux have indications of the flare only after 23:00 UT. This means that decreasing overlying field might also have played a role in the onset of the eventual eruption and the flare reconnection later adds to the acceleration of the flux rope.

\subsection{Initiation of the Eruption: Mechanism}
The observations of the eruption are plausibly compatible with the model of tether-cutting reconnection formulated by \citet{Moore1980, Moore2001}. In this model, the AR consists of an inner bipolar region with an initial sheared magnetic arcade.  When subject to converging motion towards the PIL, reconnection of adjacent opposite sheared arcade loops forms an upward rising twisted FR as proposed by \citet{Ballegooijen1989}. In Figure~\ref{Fig_schem}, we deploy this scenario of reconnection in the context of our observations. In the left panel, we add schematically on top of the HMI magnetogram the magnetic connectivities of the rising loops L1 (blue) and L2 (orange) as seen in AIA 131~\AA\ images (Earth view). For a clearer picture, we also display these loops on the HMI synoptic magnetogram as if seen from above. These loops appear as two lobes of a sigmoid as observed in other events \citep[e.g.][]{Vemareddy2018} in coherent with the observed J-shaped ribbons. This configuration is typically the precursor structure of the CME eruptions.  

The footpoint locations of the loops L1 and L2 show that they are in a highly sheared magnetic configuration above the PIL. After the tether-cutting reconnection of L1 and L2, the formed hot channel (FR in models) rises due to self or hoop force. In a successful eruption, as this one, at some point of the evolution the FR becomes unstable.  This is traced by the exponential behavior of the upward velocity.
Moreover, further reconnection adds more flux to strengthen the FR such that it can further overcome the restraining force and accelerate upwards \citep{Aulanier2010, Janvier2015}. In this process, the reconnection behind the erupting structure has the importance to further decrease the downward tension of the overlying arcade so that a positive feedback occurs on the upward motion \citep{Welsch2018}.

The location of present event, close to the solar limb, implies that a magnetic field extrapolation of the coronal field is delicate.  We still tried to see if the start of the exponential behaviour was reached when the critical index of the torus instability was reached \citep{Lin2000,Kliem2006}. The results of the potential field extrapolation, while compatible with the torus instability, are not so reliable as they involve a magnetogram taken more than one day later, so we omit to report them here. Moreover, the synchronisation of the hot channel dynamics with the EUV and X rays fluxes (Figure~\ref{fig_aia_ht}) rather points to a driving mechanism of run-away tether-cutting reconnection with a positive feedback between the upward FR motion and reconnection rate below it \citep{Moore2001,JiangChaowei2021}.

Importantly, we notice a region of remote brightening well before the onset of the eruption i.e. 22:30 UT, which may be the signature of  trigger of the onset of slow rise motion (Figure~\ref{Fig_diff}d).  Careful inspection of the animation delineates that the brightening is associated to low lying closed loop adjacent to the large scale overlying loops. The external reconnection at this remote brightening region is suggested to initiate the rise motion of the erupting feature (HFR) by removing/reducing the overarching loops. This reconnection scenario is also sketched in Figure~\ref{Fig_schem}b. The low lying loop, drawn in red, has the morphology of brightened loop seen early on (see the movie). We interpret it as the product of the reconnection of the loops in the adjacent bipolar region with the large overlying loops (both drawn in blue). The motion of the small bipolar region towards the AR positive polarity would be the key, as can be noticed in on-disk observations, to induce this reconnection co-spatial with the remote-brightening.


Finally, this event is similar to the on-disk eruptions of X-ray or EUV sigmoids or H$_\alpha$ filaments where the reconnection scenario could be ambiguous due to projection effects. Especially the on-disk observations of sigmoid event studied by \citet{Vemareddy2014} have comparable features than this limb event. 
Owing to a cool temperature background, the sigmoid contrast from the background plasma emission was very good in the hot EUV channels, which is not the case here. Observational evidences for tether-cutting reconnection have also been found by employing multitude high-resolution imaging data \citep{Vemareddy2017, ChenHechao2018}.

 
\subsection{CME Kinematics}
\label{Kinematics}

In order to derive quantitative information on the mechanisms involved in the eruption we quantify below the kinematics of the LE and the FR. In particular the temporal behavior of their velocity and acceleration provides information on the net force that accelerates the plasma.

The kinematics of this event is studied by manually tracking the CME LE and core in LASCO/C2, C3 FOV. Difference images are used to enhance the contrast. In Figure~\ref{Fig_ht_vel}, we plot the height-time observations of the core center (bright part) and leading edge of the CME observed in LASCO white light images. Height-time observations (Figure~\ref{fig_aia_ht}b) of the hot-channel observed in AIA 131~\AA~are also included. A possible manual error of 4 pixels (2.4, 48, 224 arcsec) in each data is shown.  

Within LASCO FOV, a second order polynomial is sufficient to fit well the height-time LE of the CME, as shown in the upper panel of Figure~\ref{Fig_ht_vel} (red, blue). Next, since the CME core is associated to the HFR, we fit the combined height-time data of AIA HFR and CME core with a fifth-order polynomial in order to include the large variation of the velocity. Although, the fit of the height represents well the data, computing its time derivative shows oscillatory-like behaviour, a typical behavior of fits with high order polynomial. In the early phase of rising motion this even implies negative velocities. This shows the limits of such approach to incorporate with the same analytical formula all the erupting phases and needs a focused study. Then, as for the LE, we also perform a second order polynomial fit of the CME core. 

From the above fitting to the data, we derive the velocity and acceleration information. Before the instability sets in, the FR rises with a nearly constant velocity of 40 \kms which corresponds to an acceleration of 6 \mss. Thereafter, the FR runs into a rapid acceleration phase reaching to 1400 \mss\ in the AIA FOV. This rapid acceleration phase is well within 3 $\Rsun$ as reported earlier. In LASCO FOV, the velocity of the LE (core) still increases from 1090 to 1220 (724 to 786) \kms with a lower and steady acceleration of about 8.3 (4.6) \mss. Therefore, in the coronagraph FOV, the CME continued to accelerate weakly as the case of 2015 May 9 CME studied in \citet{Vemareddy2017}. In summary, the kinematic study shows three stages of acceleration, slow, fast, and slow consistent with previous studies \citep{ZhangJie2001, Gopalswamy2003}.

\begin{figure*}[!ht] 
	\centering
	\includegraphics[width=.9\textwidth,clip=]{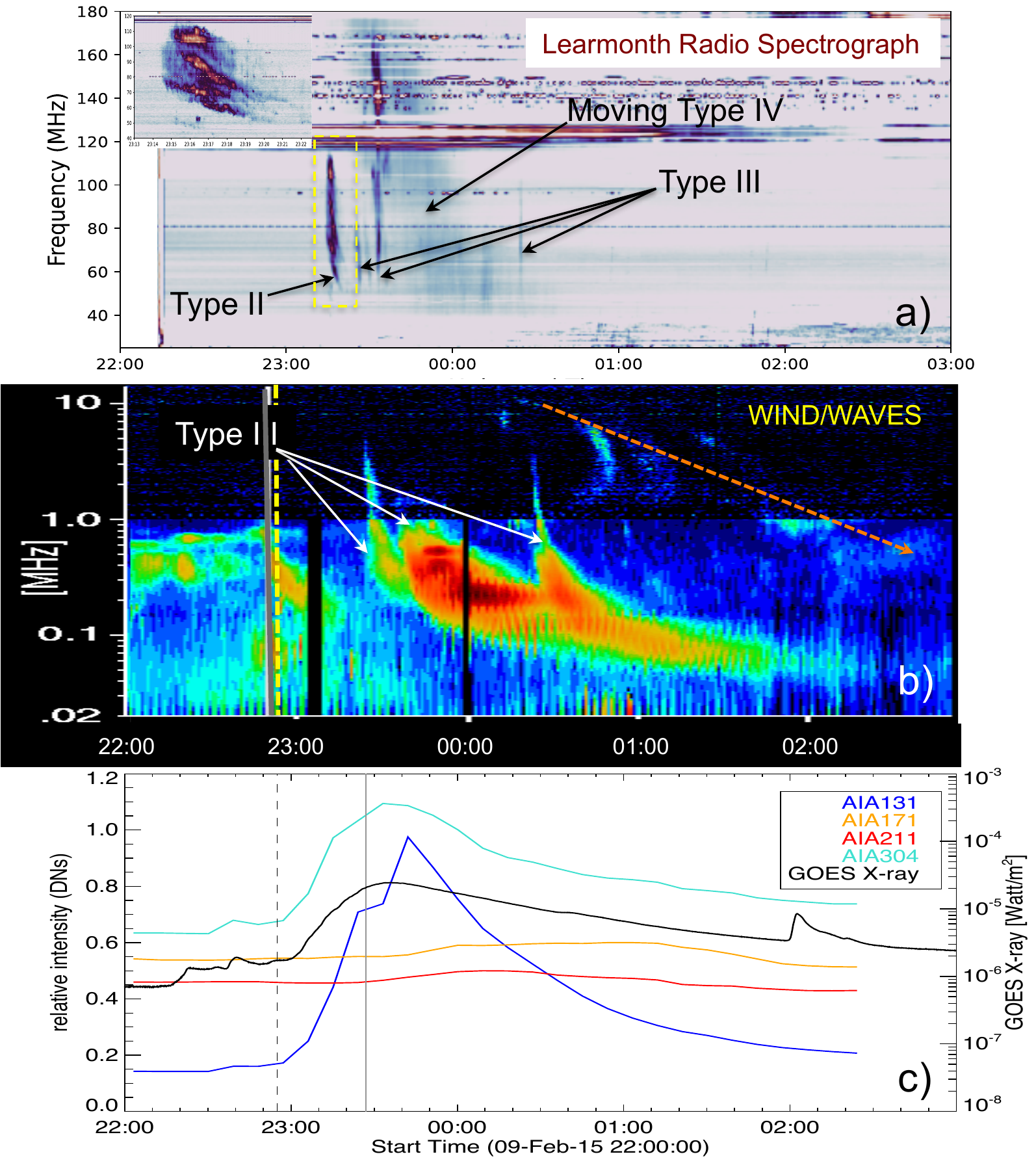}
	\caption{ Radio observations and coronal emission light curves corresponding to the CME eruption. 
	{\bf (a)} Radio spectrum observations obtained using Learmonth radio spectrometer. Arrows point to the type II and the moving type IV bursts triggered by the eruption. Inset plot is the zoomed in portion of the Type II burst outlined in the yellow rectangular box.
	{\bf (b)} Radio dynamic spectrum obtained with WIND/WAVES instrument located at the Lagrange L1 point. Arrows refer to the type III bursts originating from the CME associated flare. 
	{\bf (c)} Light-curves of the AR 12282 in EUV and X-ray wavelengths. GOES X-ray flux is integrated over the entire disk whereas EUV 131, 171, 211, 304~\AA\ channels are integrated over the AR. 
	The X-ray flux evolution is nearly co-temporal with the AIA 131 and 304~\AA\ channel fluxes.  Post flare emission corresponds to late phase enhancement in AIA 211 and 171~\AA\ wavelengths.  Dashed, solid vertical lines refer to the eruption onset (22:55 UT) and flare impulsive phase peak (23:27 UT), respectively. }
	\label{Fig_lc}
\end{figure*}


\begin{table}
	\centering
	\caption{Timeline of the eruption}
	\begin{tabular}{ l  l }
		Time [UT]   & description   \\
		\hline
		22:30 & remote brightening (Figure~\ref{Fig_diff}d)  \\
		22:42 & observed initiation of the rise motion (Figure~\ref{fig_aia_ht}\\
		22:55 & critical time of the rise motion (Figure~\ref{fig_aia_ht}) \\
		23:00 & flare start time in GOES X-ray flux (Figure~\ref{Fig_lc}c)  \\
		23:27 & peak of the flare impulsive phase (Figure~\ref{Fig_lc}c)  \\
		23:35 & flare peak time in GOES X-ray flux (Figure~\ref{Fig_lc}c  \\
		23:14 & Type-II radio burst (Figure~\ref{Fig_lc}a)\\
		23:18 & Type-IVm radio burst (Figure~\ref{Fig_lc}a)\\
		23:27 & Type-III radio burst (Figure~\ref{Fig_lc}b)\\	
		\hline
	\end{tabular}
	\label{Tab1}
\end{table}

\subsection{EUV Light Curves and Radio Spectrum}
\label{Radio}

The EUV and radio observations provide complementary information on the eruption development. Below, we complement Section~\ref{Onset} with the EUV light curves which summarize the global results of this subsection and provide a context to interpret the radio observations. These last ones allow to trace the various accelerations of electrons during the eruption: in the flare region, at the CME shock, within the erupting FR, as well as those injected along open field lines in the interplanetary space.

Figure~\ref{Fig_lc}c shows the light curves of the eruption event in different wavelengths. GOES X-ray flux is integrated over the entire disk whereas EUV 131, 171, 211, 304~\AA\ channels are integrated over the AR including the corona above the limb. The bump in AIA 304 at 22:42 UT corresponds to soft X-ray flux, which is related to remote brightening mentioned earlier. The flare start time is 23:00 UT as the light curves in EUV channels correspond well with the soft X-ray flux. Time line of this eruption event is given in the Table~\ref{Tab1}. 

The pre-flare, or initiation, phase is associated with the rise motion of the coronal loop structures L1 and L2 (Section~\ref{Onset}).
The soft X-ray peak at 23:35 UT is co-temporal with the AIA 131 and  304~\AA\ channel fluxes.  This defines the M2.4 class flare. The 131~\AA\ channel is double peaked around $5.5\times 10^4$ and $10^7$ K \citep{Boerner2012}, then it responds to the high energy deposit similarly to soft X-rays. Also, part of the magnetic energy released by reconnection is transported towards the chromosphere where it heats up the plasma.  Indeed, the associated flare ribbons are imaged in the emission of 304~\AA\ channel as this passband corresponds to chromosphere and transition region emissions. The emitted flux from the post reconnection loops falls in AIA 171 and 211~\AA\ channels which response functions are peaked around $1$ and $2 \times 10^6$ K, respectively. Then, their light curves are enhanced significantly much later than the flare peak time.   

The dynamic radio spectrum of Figure~\ref{Fig_lc}a shows a type II burst in between 23:14 UT and 23:17 UT in the frequency range 55  - 80 MHz (see the inset plot for a clear picture).  Since it is widely accepted that emission mechanism of type II bursts is plasma emission with the emitting frequency scaling as the square root of the plasma density, $N$, the height of the emission could be estimated \citep{Man1995, Gop2006}. In present data, we do not have access to the plasma density then we use a Newkirk density model adapted to ARs \citep{Newkirk1961}.  The estimated height ranges corresponding to those frequencies to be $\approx 1.45 - 1.63\, \Rsun$. This range of height is comparable with the height of the CME leading edge ($<2 \Rsun$) during the same time period (Figure~\ref{Fig_ht_vel}a). Moreover the shock is probably located at the leading edge of the CME seen in LASCO C2 and C3 (Figure~\ref{Fig_aia_CME}). Furthermore, since the source region of this event is an AR located near the limb, it is expected that only the harmonic component is observed in the dynamic spectrogram while the fundamental emission not reached Earth because of its higher directivity \citep{Thejappa2007, Bonnin2008, Thejappa2012, Ramesh2012, Sasikumar2013}.

Although, we do not have the calibrated flux densities, to compensate the gain variation across different frequency channels we have subtracted the median of the time-series separately for all channels and measured the spectral index ($\alpha$) using,

\begin{equation}
   \label{eq_spec_ind}
   \alpha = {{ln(S_1) - ln(S_2)} \over {ln(f_1)-ln(f_2)}} \,
\end{equation}
where $S_1 = 29$, $S_2 = 13$ counts are the measured amplitudes of Type II at the given frequencies $f_1 = 92$ MHz and $f_2 = 71$ MHz respectively. So, the measured spectral index of type II burst is $\approx 3$. 

The observed drift of frequency is due to the decrease of plasma density with height.  We compute this drift in the time range $t_1 =$ 23:14 and $t_2=$23:17 UT. The corresponding emission frequencies of the type II are $f_1=80$ and $f_2 = 55$  MHz.  The observed drift rate is $\Delta f / \Delta t$ where $\Delta f = f_1 -f_2$  and $\Delta t = t_2 - t_1$.  The estimated drift rate is $\approx 0.14$~MHz~s$^{-1}$. 
Then, we estimated the average Type II speed using the relation 
$v_{type II}={2 L \over f} \, {\Delta f \over \Delta t}$, where, $L=({1 \over N} {dN \over dR})^{-1}$ is the plasma density scale height ($\approx 2.31 \times 10^5$ km) and $f$ is the averaged frequency \citep{Gopalswamy2011b, Mann2005}. The estimated type II speed is $\approx 962$ \kms which is compatible with the mean speed of the CME leading edge during 23:14-23:21 UT of about 1100 \kms (see Figure~\ref{Fig_ht_vel}b).

Type III bursts are present in the dynamic radio spectrum of Figure~\ref{Fig_lc}b in the frequency range 20 KHz - 14 MHz has the meter wave counterparts (110 MHz and below) in Figure~\ref{Fig_lc}a. The Type III burst in this case is co-temporal with the X-ray and EUV emissions commencing at the peak flare time (23:27 UT) defined by time-derivative of soft X-ray flux. However, it is worth mentioning that all the three indicated type III bursts are originated after the type II burst is seen and hence presumably they are associated with the post eruption open magnetic loops. These emission correspond to electrons accelerated to a fraction of light speed along open magnetic field lines \citep{Ginzburg1958, Zheleznyakov1970, Melrose1980, Sasikumar2013, Reid2014, Mahender2020}. They could come from the reconnection of the erupting field configuration with the surrounding open field or from the flare reconnection provide a channel to open field lines \citep[e.g.][]{Masson2019}.

The dynamic spectrum also detected a moving type IV (type IVm) burst starting at 23:18 UT which lasted till 00:34 UT on February 10, 2015 (Figure~\ref{Fig_lc}a).  A type IVm burst is a broad-band continuum emission with a clear frequency drift with time. This type IVm was observed by Learmonth observatory, Australia as well as HIRAS spectrometer, Japan \footnote{https://sunbase.nict.go.jp/solar/denpa/hirasDB/2015/02/150210a.gif}. From both dynamic spectrums, the type IV is observed in the frequency range 44 - 438 MHz. The estimated drift rate is $\approx 0.09$ MHz\,s$^{-1}$. This indicates that source location moves as the CME propagates radially outward \citep{McLean1985, Gergely1986}. The height-time data shown in Figure \ref{Fig_ht_vel}a indicate that at the start time of type IVm (23:18 UT) the LE (core) of the CME was located a height of 2.1 (1.5) $\Rsun$. Similarly at the end time (00:34 UT) of type IV burst, it was at 9.4 (6.1 ) $\Rsun$. Type IVm bursts were previously associated with the CME core 
\citep{Sasikumar2014,Vasanth2019}. Although we do not have imaging observations on this day, present event could be interpreted in this general framework.


Emission mechanisms of type IV bursts are still debated and different authors suggested plasma emission, or gyro-synchrotron, or electron cyclotron maser emission as the main emission mechanism \citep[e.g.][]{Sasikumar2014, Vasanth2016, Carley2017}. Therefore, in order to investigate the emission mechanism of this particular event, we have measured the spectral index (Equation ~\ref{eq_spec_ind}). We use $f_1=60$ and $f_2=175$ MHz which correspond to $S_1=7$ and $S_2=5$ counts, respectively.
We deduce the spectral index of $\approx -0.31$ suggesting that this type IVm has a gyro-synchrotron emission mechanism \citep{Sasikumar2014}. More precisely, the negative sign of the spectral index is characteristic of a non-thermal emission mechanism. The small value of spectral index indicates an optically thin gyro-sychrotron emission. Furthermore, if the emission mechanism is due to the plasma emission, the spectral index shall be $<-3$ \citep{Melrose1975, Sasikumar2014}.  Then, since the type IVm burst has not a plasma emission origin, we are unable to estimate its velocity. 
Finally, it is worth to point that in WIND/WAVES, both type II and IVm bursts are not seen which may be due to the sensitivity of the instrument and/or the electron energy might have dropped down and then it is no longer sufficient to generate radio emission. However fragmentation that are seen along the red dotted line shown in Figure \ref{Fig_lc}b presumably are the deca-hectometric and kilometric type II burst. 


\section{Summary and Discussion}
\label{Summ}


We analysed a unique eruption of a hot plasma channel from its early evolution within the core of an AR. This eruption leads to a well observed CME from the source AR 12282 near the eastern solar limb. The ejected plasma is visible mainly within the EUV hot channel of AIA 131~\AA. An EUV brightening was first observed in the trailing part of the AR with converging motion of opposite polarities. This brightening is consistent with a middle increase in soft X-ray flux starting at 22:30 UT. At the beginning of the eruption (22:42 UT), two highly sheared loops are visible in EUV above the main AR PIL.  
This pair of loops have two of their footpoints nearby from each other on both sides of the PIL. These two loops expand, then reconnect to form a single structure, an ejected hot structure, identified as an MFR, and compact flare loops underneath. 

This eruption event is similar to the one described in the tether-cutting reconnection model \citep{Moore2001}, which is responsible for the flux rope formation and its slow rise motion. This event has comparable observational signatures than the ones expected in numerical simulations where a sheared arcade is forced to reconnect by the converging motions at the PIL \citep{Aulanier2010, Zuccarello2014}. Recent numerical simulations argue that the tether-cutting reconnection has the fundamental importance in the initiation of solar eruptions in a very simple bipolar active region \citep{JiangChaowei2021}. The observed flare ribbons are J-shaped, which is an indication that the hot-channel is a FR. This event is a unique event in which the flare ribbons is very clearly observed along with the erupting hot channel, which strongly supports that the hooked part of J-shaped flare ribbons outlines the boundary of the erupting flux rope. 
 

The kinematic study reveals a three phases evolution of the CME eruption as typically observed \citep[e.g.][]{ZhangJie2001}.  In the initiation phase, the FR slowly rises at nearly constant speed of 40 \kms.  This is followed by the flare impulsive phase from 23:00 UT where a large acceleration is present to a maximum of 1400 \mss\ within AIA FOV. This phase has an exponential rise motion of the FR, which  characterizes the development of an instability. Finally, the acceleration decreases in the propagation phase within LASCO FOV. Within C3 FOV, the CME LE (core) continues to propagate at a steady acceleration of about 8 (4) \mss. 

Since the flare impulsive phase (from 23:00 UT - 23:35 UT) observed in EUV synchronizes with the rapid acceleration of the erupting FR, the flare reconnection played a major role in the outward acceleration process within 3 $\Rsun$, similar to earlier studies \citep{HQSong2013, Vrsnak2016}. This positive feedback of reconnection on the acceleration is due to the further build up of the FR where stabilising overlying arcades are transformed to the external layer of the erupting FR \citep{Welsch2018}.

The CME eruption launched radio bursts of type-II, III and IVm as observed by ground and space based instruments. The type III bursts are identified with the emission drifting in the range 50 KHz - 180 MHz synchronous with the flare peak time for the strongest bursts.  The average type-II speed is in agreement with the speed of the leading edge of the CME. The type II burst is followed by a type IVm, typically interpreted by radio emission within the core of the CME \citep{Vasanth2016, Vasanth2019} which in this case is also the identified hot-channel FR. The start and stop times of type IVm corresponds to the CME core height of 1.5 and 6.1 $\Rsun$ respectively. Also the spectral index is negative suggesting the non-thermal electrons trapped in the closed loop structure. This event adds to the previous reports that type IVm bursts are associated with CME core being the hot-channel FR \citep{Vasanth2016, Vasanth2019}.  However, this study lacks radio imaging observations to reveal more information on the links of type IVm burst with the erupting CME structure at the source. 

Triggering and acceleration mechanisms are key points for the propagation of the CMEs in the outer corona and then in the heliosphere. For example, \citet{ZhangJie2001} suggested that the final velocity of a CME is dependent on the acceleration magnitude as well as acceleration duration, both of which can vary significantly from event to event. Continuous observations from the precursor features, especially near the limb, to the acceleration and propagation phases, so up to the CME stage in white light, are key observations to better constrain our understanding of these large ejections of plasma and magnetic field. 
We anticipate such observations from the \textit{Visible Emission Line Coronagraph} (VELC; \citealt{Raghavendra2017}) on board the upcoming Aditya L1 mission. VELC and meter wavelength radio observations probe the same heliocentric distances. Then, their simultaneous observations will give more insights on eruptions and their association with the solar radio bursts specifically type II and type IVm bursts.

\acknowledgements We thank the referees for providing very precise comments and suggestions which improved the clarity of the paper significantly. SDO is a mission of NASA's Living With a Star Program. SOHO is a project of international cooperation between ESA and NASA. We recognize the collaborative and open nature of knowledge creation and dissemination, under the control of the academic community as expressed by Camille No\^{u}s at \url{http://www.cogitamus.fr/indexen.html}. 

\bibliographystyle{apj}

\end{document}